\documentclass[aps,prd,twocolumn,nofootinbib,superscriptaddress,amssymb]{revtex4}
\usepackage{color}
\usepackage{epsfig}
\usepackage{graphicx}
\usepackage{epstopdf}
\usepackage{mathtools}
\usepackage{bbold}
\def\be{\begin{equation}}
\def\ee{\end{equation}}

\begin{document}
\addtolength{\baselineskip}{.1mm}

%\preprint{UCD-HEP-???}
\title{Black Hole Information as Topological Qubits}

\author{Erik Verlinde}

       %\thanks{}
%\altaffiliation  {}
\affiliation{Institute for Theoretical Physics, University of Amsterdam, Amsterdam, The Netherlands}

\def\spc{\hspace{.5pt}}

\author{Herman Verlinde}
\affiliation{Department of Physics, Princeton University, Princeton, NJ 08544, USA}

\date{\today}
%\date{November 19, 2007}

\begin{abstract}
The principle of balanced holography, introduced in \cite{vv1}, posits that black hole information is stored in non-local correlations between the interior and exterior. Based on this concept, we propose that black hole information decomposes into elementary units in the form of topological qubits, and is 
protected from local sources of decoherence. The topological protection mechanism ensures that  the horizon of an evaporating black hole stays young and smooth.

\end{abstract}

\def\be{\begin{equation}}
\def\ee{\end{equation}}

%\begin{document}

%\keywords{}

\maketitle
\def\mathbi#1{\textbf{\em #1}} 
\def\som{{ \textit{\textbf s}}} 
\def\tom{{ \textit{\textbf t}}} 
\def\nom{n} %{{ \textit{\textbf n}}} %{\mbox{\fontsize{13pt}{.15pt}${ \smpc \mathbf{n}\smpc }$}}}
\def\mom{m} %{ \textit{\tex%mbox{\fontsize{9.5pt}{.1pt}$\mathbf m$}}}
%\def\kom{{\textbf{\em k}}}
%_{{}%_{\mbox{\fontsize{4pt}{.5pt}{\omega}}}}}
%\def\mom{{\textbf{\em m}}}
%_{{}_{\mathtt{\omega'}}}}}
\def\la{\langle}
\def\bea{\begin{eqnarray}}
\def\eea{\end{eqnarray}}
\def\is{\! & \! = \! & \!}
\def\ra{\rangle}
\def\half{{\textstyle{\frac 12}}}
\def\cL{{\cal L}}
\def\halfi{{\textstyle{\frac i 2}}}

\def\ba{\begin{eqnarray}}
\def\ea{\end{eqnarray}}
%\section{ \label{sec:}}

%\section{}

%\subsubsection{}

%\section{Introduction}

%\subsubsection{Experimental results}

\def\ibar{\spc f\spc} %{{\, {\overline{i}}\, }}
\def\kbar{{\spc {\bar{k}\spc }}}

\newcommand{\rep}[1]{\mathbf{#1}}
\newcommand{\Tr}{\, {\rm Tr}}
\def\uU{\bf U}
\def\be{\bea}
\def\ee{\eea}
\def\delbar{\overline{\partial}}
\newcommand{\smpc}{\hspace{.5pt}}
\def\ra{\bigr\rangle}
\def\la{\bigl\langle}
\def\ccdot{\!\spc\cdot\!\spc}
\def\nspc{\!\spc\smpc}
\def\tr{{\rm tr}}
\def\bh{{\mbox{\fontsize{7pt}{.7pt}{$BH$}}}}%\mbox{\tiny \it BH}}}}

\def\Aa{{\mbox{\scriptsize \smpc \sc a}}}
\def\bfC{\mbox{{\textbf C}}}
\def\bC{\alpha} %{{\mbox{\small {\smpc \sc c}}}}
\def\nonu{\nonumber}
\def\sC{{\mbox{\scriptsize {\smpc \sc sc}}}}
\addtolength{\baselineskip}{.3mm}
\addtolength{\parskip}{.3mm}
\renewcommand\Large{\fontsize{15.5}{16}\selectfont}
\def\ra{\bigr\rangle}
\def\la{\bigl\langle}
\def\li{\bigl|\spc}
\def\ri{\bigr |\spc}

\def\hf{\textstyle \frac 1 2}

\def\cE{{\mbox{\tiny \nspc $\callE$}}}
\def\Ee{{\mbox{\scriptsize \smpc \sc e}}}
\def\Bb{{\raisebox{-.2pt}{\scriptsize \smpc \sc b}}}
\def\Zz{{\raisebox{-.2pt}{\scriptsize \smpc \sc z}}}

\def\Bbt{{\raisebox{-.2pt}{\scriptsize \smpc $\tilde{\mbox{\sc b}}$}}}
\def\Hh{{\mbox{\scriptsize \smpc \sc h}}}
\def\Aa{{\mbox{\scriptsize \smpc \sc a}}}

\def\Rr{{\mbox{\scriptsize \smpc \sc r}}}
\def\AB{{\mbox{\scriptsize \smpc \sc ab}}}

\def\BH{{\mbox{\scriptsize \smpc \sc bh}}}

\def\AE{{\mbox{\scriptsize \smpc \sc ae}}}

\def\callE{{\mbox{E}}}
\def\callBH{{\mbox{BH}}}

\def\callH{{\mbox{H}}}
\def\callB{{\mbox{B}}}
\setcounter{tocdepth}{2}
%\tableofcontents

%\setlength{\parskip}{1mm}
\subsection{Introduction} 
\vspace{-3mm}

Quantum information can not always be localized.  In condensed matter physics, this fact underlies the physical~realization of topological qubits and topological order~\cite{kitaev,tops,wen}. 
The simplest example are the famous majorana qubits  \cite{kitaev}.
The majorana algebra $\gamma_1^2 \!=\! \gamma_2^2  \!=\!1$, $\{\gamma_1,\gamma_2\}\!=\!0$ can be realized on a single qubit, via the identification with Pauli operators $\gamma_i=\sigma_i$.
In case the majorana operators $\gamma_i$  are associated with quasi-particles placed at different locations \cite{kitaev}, the  qubit  is non-locally stored and therefore `topological protected' in sense that %the size of its Hilbert space and 
its quantum purity is insensitive to local perturbations.

Although there are many parallels, a black hole is not a condensed matter system. The concept of non-local storage
and topological protection of quantum information, however, is likely to be equally relevant and revealing in both contexts.

The black hole information  and firewall paradox \cite{bh,infoloss,mathur, amps, ampss} are quantum information theoretic contradictions following from certain locality assumptions. We introduce the following three space-time regions
\vspace{1.5mm}

${}$~~~~\parbox{7.6cm}{
H\, :~~black hole interior or stretched horizon~~~~~~~~~~~~${}$
\\[.7mm]
E\, :~~entanglement zone, region close to the horizon\\[.7mm]
R\, :~~Hawking radiation, region far outside  horizon}

\vspace{1.5mm}

\noindent
It is often assumed  that quantum information is built up from local units (qubits)  situated in one of these space-time regions \cite{amps,ampss,hp,bitmodels}. This hypothesis indeed holds for 
most local quantum field theories.

However, as string theory and the information paradox itself demonstrate, the
medium by which black holes store and release information can not be fully captured
by local QFT. Any plausible mechanism for information retrieval must involve hidden long range 
quantum correlations.  Holographic realizations of black hole space-times, like the gauge gravity duality, indicate 
that these correlations are coded in the structure of space itself: space represents a highly entangled quantum state, and 
its uniformity and locality are emergent properties, rather than fundamental principles of the underlying microphysics.

The principle of balanced holography introduced in \cite{vv1} posits  that black hole information is stored in non-local correlations  between microscopic interior and exterior degrees of freedom.  Here we will show that this information is naturally organized in a form which is mathematically and physically similar to topological
qubits. Furthermore, based on this similarity and general arguments, we propose that it is protected from decoherence by local probes. 
 In \cite{vv1} it is shown how for a balanced black hole state, one can remove the firewall via a universal unitary entanglement swap. 
 The aim of this note is to show that this result is  stable under time evolution.

Following \cite{vv1}, we introduce two types of qubits: virtual qubits,  whose state is fixed by microphysical vacuum conditions, and real qubits, which carry the quantum~information of the freely chosen initial state. In total, 
we distinguish four forms of quantum information:\\[1.5mm]
${}$~~\parbox{8.3cm}{\addtolength{\baselineskip}{.5mm}
-\, information hidden inside the black hole region H\\
-\,  hidden microscopic information in the zone region E\\
-\, virtual QFT modes inside the zone region E\\
-\, real Hawking radiation in the far away region R}

\vspace{1mm}

\noindent
In our terminology,
 quantum field theory modes inside the zone are virtual qubits, and outside the zone they are 
real qubits. Note, however, that the categories of hidden quantum information are listed  as distinct from
the latter two. Black hole evaporation involves the transfer of hidden to visible quantum information. 
The detailed transfer mechanism is still largely unknown. In string theory, it involves
the holographic dictionary between microphysical D-brane  degrees of freedom (c.f. \cite{fuzzball}) and effective field theory variables in the
emergent near horizon geometry. In semi-classical terminology, it involves virtual pair creation and a non-local tunneling process from  the 
black hole interior to the region outside the zone.

\vspace{-3mm}

\subsection{Topological qubits I}

\vspace{-3mm}

What is a topological qubit? 
Consider four majorana operators $\gamma_i$ defined by the algebra
$\{ \gamma_i, \gamma_j\}\!=\!2 \delta_{ij}$.
%\be
%\lbrace\gamma_i,\gamma_j\rbrace =2 \delta_{ij}.
%\ee
The $\gamma_i$ can be represented on a Hilbert space of two qubits. 
The 2-dimensional `code subspace' specified~by
\be
\label{special} \gamma_5 \li \Psi \ra = \li \Psi\ra~~~
{\rm with}~~~\gamma_5=\gamma_1\gamma_2\gamma_3\gamma_4,
\ee
defines a single `topological qubit'  \cite{tops}. This terminology stems from the fact that, in case $\gamma_i$ represent distant localized majorana excitations,
the quantum information carried by the topological qubit is stored non-locally and hence protected from local sources of decoherence.
This protection is a characteristic of the underlying topologically ordered many body state \cite{wen, tops, kitaev}.
Physical operators, that commute with the constraint (\ref{special}) and change the topological qubit, are necessarily bi-local.

The four  majoranas can be paired up into two Dirac oscillators via $c^\dag_{ij} =( \gamma_i - i \gamma_j)/\sqrt{2}$. 
States in the code subspace have an even number of Dirac fermions
\be \label{tops}
 \li \Psi\ra =  \alpha_0\spc %| 0 \rangle_{\rm top}\! =\nspc 
\li 0_{12} 0_{34} \ra + \alpha_1\spc %| 1 \rangle_{\rm top} \! = 
\nspc \li 1_{12} 1_{34} \ra,
\ee
with $n_{ij}$ the Dirac fermion number  and $\alpha_{0}$  and $\alpha_1$ some arbitrary complex amplitudes. 
%The topological qubit is maximally entangled if $|\alpha_{0}|^2 = |\alpha_1|^2 = 1/2$

A topological qubit carries one qubit of free quantum information. We call this the `real' or `logical' qubit. 
The other qubit is prescribed to be in a fixed state by the physical state condition (\ref{special}). This `virtual qubit' represents confined quantum information.
In proposed physical realizations of topological qubits \cite{tops}, the ground state of the virtual qubit in fact represents a many body ground state of a topologically ordered medium, 
that serves to protect the logical qubit from decoherence \cite{kitaev}.

The decomposition of a topological qubit into a virtual and logical qubit can be made explicit by applying a CNOT operation -- flipping the first qubit, provided the second qubit reads out as 1 --
to the two qubit state (\ref{tops}), and writing it as $|\Psi_0 \rangle = {\bf U}_{{\! }_{\rm CNOT}} |\Psi\rangle$,
where $|\Psi_0\rangle = |0_{12}\rangle\; ( \alpha_0 | 0_{34} \rangle + \alpha_1|1_{34}\rangle)$.

A topological qubit constitutes a smart form of entanglement: it is neither a random Bell pair, nor a unique entangled state of two qubits. It is balanced right between the two:
it carries coherent quantum information, and just enough organized free space to move it around. This property will be important in what follows.

\vspace{-3mm}

\subsection{Topological Protection of BH Information}

\vspace{-3mm}

In \cite{vv1}, we introduced the principle of balance holography, which implies that a general state of a young black hole and its environment
can be written in the form
\be
\label{gnstate}
\li \mathbf{\Psi} \ra = \sum_{i} \,  \alpha_{i}\, \li \spc i \spc \ra_{\! \Hh}\, \li \spc i \spc \ra_{\! \cE}\, .
\ee
Here the sum runs over all the $e^{S_\BH} = 2^N$ internal states of the black hole. H denotes the interior, and 
E denotes the entangled environment of the young black hole.  Since the young black hole has not
yet produced an appreciable amount of Hawking radiation, E predominantly consists of virtual quantum field theory modes and  hidden microscopic degrees of freedom \cite{fuzzball} situated inside the `zone', the  region between the horizon and the centrifugal barrier.
The black hole quantum information is stored in $2^N$ independent complex amplitudes $\alpha_i$, and this represents $N$ qubits of real quantum information.

Here we propose that the quantum information stored in H and E
is protected from local disturbances. We adopt this as a postulate:

\medskip

\parbox{8.2cm}{
\it Black hole information is protected from local sources of decoherence. As a result, it can not be measured, altered, or mined 
by local probes inside the zone. 
  }
\medskip

Many body quantum systems in which the ground state sector enjoys this code property are said to be `topologically ordered' \cite{wen}.
Physically, the `topological' protection  arises because the energy splitting between the ground states is exceedingly 
small compared to the minimal energy jump induced by local probes. 
This applies to our setting. Black hole micro-states have an enormous level density. So if we fix the total energy of the quantum state to lie within 
a narrow energy band $M$ and $M+ \delta M$, the black hole micro-state can only be changed  (i) actively, via a slow and careful 
non-local process inside the zone or (ii) passively, by measuring on-shell Hawking radiation after it has escaped the zone.
Virtual radiation inside the zone can only be measured by accelerating probes. Just as in Rindler space \cite{unruhwald}, such virtual modes are 
produced by the probe itself, and do not carry microscopic information about the black hole state.

\vspace{-3mm}

\subsection{Topological Qubits II}

\vspace{-3mm}

In balanced holography, black holes quantum information naturally organizes itself in elementary units, mathematically identical to 
topological qubits \cite{tops}. Indeed, it is easy to convince oneself \cite{vv1} that  
the Hilbert space of maximally entangled young black hole states  of the form (\ref{gnstate}) factorizes, to a very good approximation, into the tensor product of $N$ 
entangled qubit pairs, each spanned by states of the  form 
\be \label{xtops}
 \li \Psi\ra =  \alpha_0\spc %| 0 \rangle_{\rm top}\! =\nspc 
\li 0 \ra_{\! \Hh} \li 0 \ra_{\!\Ee} + \alpha_1\spc %| 1 \rangle_{\rm top} \! = 
\nspc \li 1 \ra_{\! \Hh} \li 1 \ra_{\! \Ee},
\ee
where  $\alpha_{0}$  and $\alpha_1$ are arbitrary complex amplitudes. 
The state (\ref{xtops}) is maximally entangled if $|\alpha_{0}|^2 = |\alpha_1|^2 = 1/2$.
Indeed, the Hilbert space spanned by $N$ topological qubits of the form (\ref{xtops}) is $2^N$ dimensional.

%, each forming a bound state of one virtual and one real qubit.

%\begin{figure}[t]
%\begin{center}
%\includegraphics[scale=.6]{topq.pdf}
%\caption{\small A topological qubit made up from 4 majoranas.}
%\end{center}
%\vspace{-0.5cm}
%\end{figure} 

%In a condensed matter context this protection is a characteristic of the underlying topologically ordered many body state. 
%Although our context is quite different, the physical concept of non-local storage and topological protection of quantum information is very relevant to our problem.

\begin{figure}[t]
\begin{center}
\includegraphics[scale=.54]{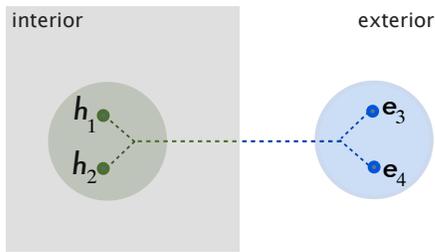}
\caption{\small Black hole information as  a topological qubit. 
The majorana pair $(h_1,h_2)$ lies in the black hole interior H, while the
pair $(e_3,e_4)$ resides in the entangled environment E. The dotted line indicates the fermion number constraint (\ref{ntops}).}
\end{center}
\vspace{-0.5cm}
\end{figure}

The majorana representation of the two qubit state (\ref{xtops}) is depicted  in fig 1. The left majorana pair $(h_1,h_2)$ are associated with degrees of freedom that live inside the black hole interior H
and the right pair $(e_3,e_4)$ are stored in degrees of freedom of the entangled environment E. 
The majorana algebra $\{h_i,h_j\} = 2 \delta_{ij}$, $\{e_i,e_j\} = 2 \delta_{ij},$
$\{ h_i , e_j \} = 0,$ can be represented on the Hilbert space of two qubits.
The two-dimensional
code subspace of the topological qubit
is selected by the stability condition~(\ref{special})
\be
\label{ntops}
& & \qquad (-1)^{n_{12}} = i h_1 h_2,\nonumber \\[-2.5mm]
n_{12} \li \Psi \ra = n_{34} \li \Psi\ra,   & & % {\rm with} \left\{ \qquad \ \right.
 \\[-2.5mm]  & & \qquad (-1)^{n_{34}} = i e_3 e_4,\nonumber
\ee 
which projects the virtual qubit on a specified state.
So $ \li \Psi \ra$ indeed takes the balanced form (\ref{xtops}) with 0,1 the binary `Dirac fermion number' defined in (\ref{ntops}).

The majorana qubits give an apt physical representation of the long distance quantum correlations, 
that carry the black hole information. The spatial separation of the majorana excitations embodies the topological 
protection mechanism, that ensures that local excitations at the horizon can not access or change the coherent information stored in the
microscopic state of the black hole. Even for an observer with access to the real Hawking radiation, the task of 
capturing and decoding the state (\ref{xtops}) of a typical entangled qubit will typically require a very long time \cite{hh}.
The majorana representation naturally incorporates both these characteristics.

As explained, the topological qubit state (\ref{xtops}) can be thought of as a bound state of a logical qubit, that carries the free black hole information,
and a virtual qubit that labels the total $\mathbb{Z}_2$ Dirac fermion number $(-1)^{n_{12} + n_{34}}$. One can think of the virtual qubit as specifying 
the delocalized spatial wave-function of the logical qubit. If the virtual qubit is in the ground state, the logical qubit has a smooth wave function
across the horizon. If it is in the excited state, the wave function involves a discontinuity in the form of a spin flip. We refer to \cite{vv1} for some 
further discussion of the role of these two types of qubits.

In condensed matter physics, topological protection relies on properties of the separating medium.
We will not attempt to give such a microscopic description here.  
Rather we turn the problem around: we postulate that the topological qubits give an accurate representation of how black hole information is stored. Our task then is to reconcile this structure with semi-classical bulk physics in the zone and exterior regions. We can view this task as assembling
a consistent holographic dictionary.

\vspace{-2mm}

\subsection{Locality and Complementarity}

\vspace{-3mm}
Consider the following  set of majorana observables
\be
\label{aliceop}
a_1 \is h_1, \qquad \ \ \ \ \, a_2 =i h_2 e_3 e_4,\nonumber\\[-2mm]\\[-2mm]
B_1 \is i h_2 e_3 , \qquad  B_2 = i h_2 e_4\, .\nonumber
\ee
These satisfy $\{a_i,a_j\} = \{B_i, B_j\} = 2\delta_{ij}$, and the following two special properties
(see eq (\ref{ntops}))
\be
\label{avac}
(a_1 + i a_2) |\Psi\rangle = \!\! &  0,& \\[2mm]
[ a_{\smpc i}, B_{\nspc\nspc j} ] = 0. & &
\ee
We will interpret eq (\ref{aliceop}) as a holographic reconstruction map between hidden microscopic units $(e_i,h_j)$ and the `bulk' variables $(a_i,B_j)$ used by low energy observers.

Consider two observers: an infalling observer (Alice),
and an observer (Bob) who stays outside the zone. Black hole complementarity \cite{comp} requires that:

\noindent
i) Alice  should experience every black hole state $|\Psi\rangle$ as a no drama state with a smooth horizon geometry.
Moreover, her operators should be capable of creating excited states above the vacuum. Yet,
according to our topological protection postulate, she should not be able to measure or alter the black hole information.

\noindent
ii) Bob, on the other hand, detects real Hawking particles. He has, at least for those particles that he can capture, 
full access and control over the active quantum information of the black hole state. 

\noindent
iii) Alice's operators must (anti-)commute with those of Bob. This guarantees that 
Bob's operators preserve the local vacuum conditions at the horizon, and that 
Alice and Bob can not communicate  by acting with their respective local observables.

Comparing these complementarity requirements with the eqs (\ref{aliceop}) and (\ref{avac}), we are naturally (i.e. independently from our suggestive use of notation)
 led to identify the $a_i$ oscillators with observables used by Alice, and the $B_i$ oscillators with the observables of Bob.
The $a_i$ oscillators  can only access the virtual qubit: they experience the state $\li \Psi\ra$ as a unique vacuum state. 
The logical qubit that stores the black hole information is hidden from Alice's view.
Bob's operators $B_i$, on the other hand,  generate all Pauli operators of the logical qubit. This fits our expectations i) though iii).
 
Indeed, we can make the formulas (\ref{aliceop}) and (\ref{avac}) look a bit less abstract, by introducing the raising and lowering operators $
{\bf a}\spc = ( a_1\! +\nspc  i a_2)/\sqrt{2}$ and ${\bf B}\spc = ( B_1 +\spc  i B_2)/{\sqrt{2}}$, which satisfy two Dirac algebras $\{{\bf a}^\dag,{\bf a}\} = \{{\bf B}^\dag, {\bf B}\} = 1$. 
Eq (\ref{avac}) then becomes a standard annihilation condition ${\bf a} \li \Psi \ra = 0$.
This suggests we can assign (some of) the ${\bf a},{\bf a}^\dag$ oscillators the physical meaning of annihilation/creation modes of an infalling Kruskal 
observer.
Similarly, annihilation/creation operators of the outside observer are assembled out of the ${\bf B}$ and ${\bf B}^\dag$ oscillators.
 The ${\bf a}$ and ${\bf B}$ oscillators commute.\footnote{The definition of the ${\bf B}$ oscillators in (\ref{aliceop}) can be changed, by replacing $h_2$ by $h_1$.
 The ${\bf a}$ and ${\bf B}$ oscillators then anti-commute. Either option would be consistent with locality. In addition,  there is a gauge freedom to  
 conjugate the oscillators with any unitary operator that commutes with the stability conditions of the topological qubit.}

There are perhaps still a few mysterious aspects to the above identifications. Why  aren't we identifying the virtual radiation mode $a_1$ in fig 2 as the radiation 
mode that eventually gets detected by an asymptotic observer?  This is exactly where black hole complementarity steps in: the asymptotic and infalling
observer both have a drastically different perspective on the near horizon region. So there is no reason 
to identify Alice's virtual modes  with those seen by the asymptotic late time observer.

How can the perspective of the two observers be so disparate?
 Alice and Bob expand their quantum fields via different mode functions, determined by their respective local geometries. The
metric is a classical mean field quantity associated  to a complex quantum mechanical medium. Modes that propagate through 
this medium get distorted, and in a complete quantum description, the resulting Bogolyubov coefficients or gray body factors are operator 
valued quantities,  which probe sensitive quantum information about the micro-state.
Hawking evaporation involves a tunneling process from virtual to real particles: it starts with virtual pair creation, yet extracts real quantum information from the black hole.
Considering these facts, it is not surprising that the transition from inside to outside the zone 
involves a transmutation from virtual to real qubits.

\begin{figure}[t]
\begin{center}
\includegraphics[scale=.5]{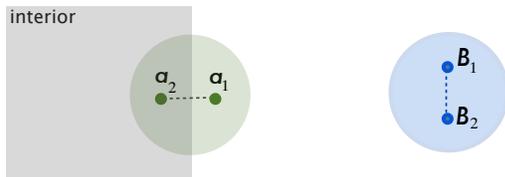}
\caption{\small After the entanglement swap the logical qubit (Bob's) and virtual qubit (Alice's) are disentangled. Bob's qubit is the Hawking radiation, that carries out the information. Alice qubit is in the ground state, and represents 
a virtual vacuum pair. It can decouple from the horizon. }
\end{center}
\vspace{-0.5cm}
\end{figure}

\vspace{-2mm}

\subsection{Time evolution}
\vspace{-3mm}

A young black hole quickly settles into a maximally mixed state 
with mostly short and medium range entanglement: it has not had time yet to generate long distance entanglement via Hawking emission.
The black hole information is all stored in hidden quantum correlations between the black interior and the zone.
The evaporation process involves the gradual release and transfer of this hidden information into effective quantum field
theory modes, measurable to a local low energy observer.

In our qubit description, the basic time step is visualized in the transition from fig 1 to fig 2. It depicts the rearrangement of majorana variables given in eqs (\ref{aliceop}).
We propose that this rearrangement represents the result of the Hawking tunneling process. This tunneling rate per qubit is extremely small.
Hence for any given qubit, the process takes place extremely slowly.
The tunneling interpretation also explains the apparent non-locality 
of the map (\ref{aliceop}). Each time step releases one unit of black hole information in the form of a logical qubit, 
represented by Bob's majorana pair $B_i$. Simultaneously, it isolates one virtual qubit, represented by the virtual majorana pair $a_i$. 
 This virtual qubit does not carry any entanglement or quantum information: it just looks like a piece of vacuum space time, that 
can simply decouple from the horizon. The horizon recedes and has released one balanced pair or qubits. So the black hole remains in a balanced state.

This time jump description is slightly misleading, however, since it assumes knowledge about whether a particular qubit has left the black hole or not. So secretly, the step
involves a projection onto a given outcome of a measurement.  This projection removes the long range entanglement between the interior  and the Hawking radiation, 
and thereby rejuvenates the black hole state. Since this entanglement is thought to be the root of the problem, we need to be a bit less cavalier about removing
it.

Let us include the entanglement with the early radiation. 
Consider an old black hole  with entropy $S_{\BH} = N \log 2$. The Hilbert space of its early radiation R is much bigger than that of the black hole interior H. 
We separately define E as the entangled environment of  a {\it young} black hole with entropy $S_{\BH} = N \log 2$. The Hilbert space of E is of the same size as that of the black hole. E is not a subsector of R, because E consists of virtual modes and hidden microscopic degrees of freedom inside the zone,
while R consists of real Hawking radiation.

We have seen above that the Hawking radiation is evenly extracted from E and H.
 The radiation qubits in R are entangled with the real qubits of the black hole state.
 For an old black hole, this entanglement has reached its saturation limit.  
We deduce that the state of the old black hole space time takes the schematic form (c.f.  \cite{nomura})
\be
\label{oldstate}
\li \Psi \ra_{\rm \nspc old} =\, \sum_i\;  \li\spc i, i\spc \ra_{{\! }{\Hh\nspc \Ee}} \, |\Phi_i\ra_{\! {\Rr}}
\ee
where $ \li\spc i, i\spc \ra_{{\! }{\Hh\nspc \Ee}}$ is short-hand of a balanced basis state (c.f. eq (\ref{gnstate})) of a {\it young} black hole, consisting of an interior H and its
entangled environment E, and $|\Phi_i\ra_{\! {\Rr}}$ denotes a state of the early radiation. The sum runs over all $2^N$ physical states of the young black hole space-time.
 The entanglement between the combined  HE region and the early radiation R saturates the B-H bound.

As a consistency check, let us show that the form of the old black hole state (\ref{oldstate}) is robust under time evolution.
Let time elapse a bit. The black hole has emitted some additional Hawking particles, in a state which we denote by $|n\rangle$. 
The time evolved state then looks as follows
\be
\label{newoldstate}
\li \Psi' \ra_{\rm \nspc old} = \sum_{i,j, n}\spc  C^{\, i}_{jn}\, \li \spc j,j\spc \ra_{{\! }{\Hh\nspc \Ee}}\, \li n, \Phi_i \ra_{\! {\Rr}} 
\ee
where $i$ and $j$ denote the local black hole state before and after the emission, and $C^{\, i}_{jn}$  is the microscopically determined emission coefficient.
In writing the time step (\ref{newoldstate}), we used that the time evolution by means of the microscopic interaction Hamiltonian preserves the stability conditions 
of the virtual qubits, and thereby the balanced holography property of the local state $| j,j\spc \rangle_{{\! }{\Hh\nspc \Ee}}$.

By combining the early plus late radiation into one new `early radiation' state  
\be
%\quad\mbox{with}\quad 
\li \Phi_j\ra_{\! {\Rr}}  = \sum_{i,n}\spc C^{\, i}_{jn}\spc \li n, \Phi_i \ra_{\! {\Rr}} 
\ee
we can write the final state (\ref{newoldstate}) in the same form as the initial state as
%\sum_i | i, i \ra |\Phi_i\ra  \to  
$\sum_j |j,j\ra_{{\! }{\Hh\nspc \Ee}} \li \Phi_j\ra_{{\! }{\Rr}} .$
In this way time evolution brings the entanglement and microscopic entropy of the black hole down by one unit in each time step:
the evaporation proceeds by emission of complete topological qubits from the local black hole environment HE into the radiation
sector $R$. So the form (\ref{oldstate}) for the old black hole state is consistent with time evolution.

A time evolution equation of the general form (\ref{newoldstate}) 
is also perfectly consistent with Page dynamics \cite{page}.  Extending the evolution rule  (\ref{newoldstate}) to black holes of all ages, and assuming that the coefficients  $C^{\, i}_{jn}$ are ergodic matrices, one can show that 
the entanglement between the balanced black hole state and the Hawking radiation follows the Page curve, as it should.

Via the formula (\ref{newoldstate}) we have made direct contact with the previous paper \cite{vv}. In \cite{vv}, a detailed study was presented of a time evolution equation of the form (\ref{newoldstate}),
based on the assumption that the transition amplitudes $C^{\, i}_{jn}$ are ergodic matrices. It was shown that, with use of some quantum error correction (QEC) technology,
one can construct the interior QFT observables, inside of the black hole horizon, provided that the state $\li \Psi\ra$ can be restricted to lie inside a suitable code subspace of the total Hilbert space. Via our balanced holography principle, we have supplied a concrete proposed identification of this code subspace, and a physical rationale for why
it is protected by the microscopic dynamics. 

The accuracy of the QEC operation and reconstruction of the interior observables is set by the codimension of the code subspace within the full Hilbert space, 
or equivalently, by the dimension of the Hilbert space of virtual qubits. This indicates that, by combining our present results with those of \cite{vv}, the interior QFT observables and operator algebra can be reconstructed with an accuracy of order $e^{-S_\BH/2}$. We leave the details of this analysis for future work.

\vspace{-3mm}

\subsection{Conclusion}

\vspace{-2mm}

Does the old black hole state (\ref{oldstate})  have a smooth horizon?
We claim that it does. In fact, the firewall problem was already solved once we were able to give a state independent definition of the operators used by an infalling observer for arbitrary balanced young black holes \cite{vv1}. Since an old black hole state can be expanded in terms of balanced young black hole states,
 our construction of the infalling operators also works for old black hole states (\ref{oldstate}). Linearity of quantum mechanics implies that, from nearby, age doesn't matter.

So what is the mistake in the AMPS reasoning? Why doesn't the long distance entanglement with the Hawking radiation crowd out the
short distance virtual entanglement? In short,
the reason is that a virtual Hawking pair is in a unique state. It can therefore  not carry any black hole information, and can not be entangled 
with anything else. The early radiation is instead entangled with the logical qubits, the hidden microscopic correlations that carry the black hole information. 
This kind of entanglement between real qubits has no local physical consequences, thanks to linearity of quantum mechanics.
It does not affect the virtual qubits responsible for smoothness of the horizon,  and the vacuum
conditions  at the horizon thus remain intact.

\begin{center}
{\bf Acknowledgement}
\end{center}
\vspace{-2mm}

\noindent
We thank  Ignacio Cirac, Daniel Harlow,  and X.G Wen for helpful discussions.
The research of E.V. is supported by the Foundation of Fundamental Research of Matter (FOM), the European Research Council (ERC), and a Spinoza grant of the Dutch Science Organization (NWO). The work of H.V. is supported by NSF grant PHY-0756966.


\begin{thebibliography}{99}

\bibitem{vv1} E. Verlinde and H. Verlinde, Passing though the Firewall,   arXiv:1306.0516
  \bibitem{kitaev} A Yu Kitaev, Annals of Physics {\bf 303} (2003) 2Ð30;   Phys.-Usp. {\bf 44}, 131 (2001)
   \bibitem{tops} 
  C.~Nayak, S.~H.~Simon, A.~Stern, M.~Freedman and S.~Das Sarma,
  %``Non-Abelian anyons and topological quantum computation,''
  Rev.\ Mod.\ Phys.\  {\bf 80}, 1083 (2008). 
  \bibitem{wen} X.-G. Wen, Adv. Phys. 44, 405 (1995). 
\bibitem{bh} S. Hawking, %Particle Creation by Black Holes, 
Commun.Math.Phys. 43 (1975) 199;   J.~D.~Bekenstein, %``Black holes and entropy,'' 
 Phys.\ Rev.\ D {\bf 7}, 2333 (1973).
\bibitem{mathur}   S. D. Mathur, %The Information paradox: A Pedagogical introduction, 
 Class.Quant.Grav. 26 (2009) 224001%, 0909.1038.
 \bibitem{infoloss}  S.~W.~Hawking,
  %``Breakdown of Predictability in Gravitational Collapse,''
  Phys.\ Rev.\ D {\bf 14}, 2460 (1976).
\bibitem{amps}A. Almheiri, D. Marolf, J. Polchinski and J. Sully, Black Holes: Complementarity or Firewalls?, 1207.3123 
\bibitem{ampss} 
  A.~Almheiri, D.~Marolf, J.~Polchinski, D.~Stanford and J.~Sully,
  %``An Apologia for Firewalls,''
  arXiv:1304.6483 [hep-th].
  \bibitem{hp} P. Hayden and J. Preskill, %Black holes as mirrors: Quantum information in random subsystems, 
JHEP 0709 (2007) 120%, 0708.4025
\bibitem{bitmodels}        S. G. Avery, 1109.2911;
S. B. Giddings and Y. Shi, 1205.4732
\bibitem{fuzzball}        S. D. Mathur, The Fuzzball proposal for black holes: An Elementary review, Fortsch.Phys. 53 (2005) 793Ð827, hep-th/0502050
\bibitem{comp} L. Susskind, L. Thorlacius and J. Uglum, %The Stretched horizon and black hole complementarity, 
Phys.Rev. D48 (1993) 3743Ð3761, hep-th/9306069;
 Y. Kiem, E. Verlinde and H. Verlinde, %Black hole horizons and complementarity, 
 Phys.Rev. D52 (1995) 7053 %, hep-th/9502074;
 \bibitem{hh} D.~Harlow and P.~Hayden, 1301.4504; 
 \bibitem{unruhwald} 
  W.~Unruh and R.~Wald,
  %``What happens when an accelerating observer detects a Rindler particle,''
  Phys.\ Rev.\ D {\bf 29}, 1047 (1984).
\bibitem{nomura} 
  Y.~Nomura and J.~Varela,
  %``A Note on (No) Firewalls: The Entropy Argument,''
1211.7033;
  %%CITATION = ARXIV:1211.7033;%%
  %6 citations counted in INSPIRE as of 30 May 2013 
 Y.~Nomura, J.~Varela and S.~J.~Weinberg,1304.0448.    
\bibitem{page}        D. N. Page, %Average entropy of a subsystem, 
Phys.Rev.Lett. 71 (1993) 1291Ð1294 %, gr-qc/9305007
; D. N. Page,  hep-th/9305040
\bibitem{vv} 
  E.~Verlinde and H.~Verlinde,
  %``Black Hole Entanglement and Quantum Error Correction,''
  arXiv:1211.6913 [hep-th].  
  \end{thebibliography}
\end{document}